\documentclass[%
preprint,
amsfonts,
amsmath,amssymb,
aps,
prl,
]{revtex4-1}

\usepackage{graphicx}
\usepackage{dcolumn}
\usepackage{bm}
\usepackage{upgreek}
\usepackage{amsmath}

\usepackage[dvipsnames]{xcolor}
\usepackage{txfonts}
\usepackage[normalem]{ulem}
\newcommand{\soutm}[1]{\ifmmode\text{\sout{\ensuremath{#1}}}\else\sout{#1}\fi}
\usepackage{hyperref}
\allowdisplaybreaks

\begin{document}

\title{Ultralow radiative heat flux by Anderson localization in quasiperiodic plasmonic chains}

\author{Yizhi \surname{Hu}$^1$}

\author{Kun \surname{Yan}$^1$}
\author{Wei-Hua \surname{Xiao}$^1$}
\author{Xiaobin \surname{Chen}$^{1,2,*}$}%
\affiliation{$^1$School of Science, State Key Laboratory on Tunable laser Technology and Ministry of Industry and Information Technology Key Lab of Micro-Nano Optoelectronic Information System, Harbin Institute of Technology, Shenzhen, Shenzhen 518055, China\\
$^2$Collaborative Innovation Center of Extreme Optics, Shanxi University, Taiyuan 030006, China\\
$^*$Email: chenxiaobin@hit.edu.cn}%

\begin{abstract}
Anderson localization, arising from wave interference in disordered systems, profoundly hinders energy transport, yet its impact on radiative heat flux in many-body thermophotonic systems remains unclear. Here, we demonstrate a three-order-of-magnitude suppression of radiative heat transfer, resulting in ultralow radiative heat transfer, in a one-dimensional quasiperiodic chain of plasmonic nanoparticles. This suppression in radiative heat transfer is directly correlated with mode localization, as revealed by the mode decomposition of the transmission coefficient, which serves as evidence of Anderson localization. Furthermore, we elucidate the dependence of radiative thermal conductance reduction on interparticle spacing and material damping rates, uncovering the interplay between intrinsic Ohmic losses, mode localization, and long-range many-body interactions. Our findings advance the understanding of wave-mediated thermal transport in disordered photonic structures and suggest strategies for tailoring nanoscale heat management via engineered disorder.

\end{abstract}
\maketitle

\section{Introduction}

Disorder, described qualitatively as irregularity in spatial patterns, is ubiquitous in both natural materials and artificial media. Disordered systems encompass a range of structural phases, including quasiperiodic, correlated, and uncorrelated disordered structures. Each of these phases is characterized by a distinct level and pattern of disorder~\cite{yu2021engineered}. A prominent phenomenon observed in disordered systems is the localization of quantum waves, a many-body coherence effect first explored by Anderson in 1958, now referred to as Anderson localization~\cite{anderson1958absence}. The theory of Anderson localization predicted a complete halt of particle, charge, spin, or energy flux due to random disorder, ultimately making conductive materials insulating. Since Anderson localization results from the interference of coherent waves, it applies not only to quantum waves but also to classical waves such as electromagnetic~\cite{de1989transverse,schwartz2007transport,yamilov2023anderson}, acoustic~\cite{condat1987resonant,arregui2019anderson,Hu2025}, elastic~\cite{hu2008localization,amir2013emergent}, and hydrodynamic waves~\cite{abraham2024anderson}. To date, both theoretical and experimental reports have consistently demonstrated that strong localization of electromagnetic waves can be achieved in random~\cite{lahini2008anderson,leonetti2014observation,shi2018strong}, correlated disordered~\cite{conley2014light}, and quasiperiodic media~\cite{lahini2009observation,levi2011disorder,jeon2017intrinsic}. From the perspective of electromagnetic applications, localized modes in disordered photonic systems offer excellent opportunities for stable random lasing~\cite{liu2014random}, broadband absorption spectrum~\cite{hsu2015broadband,yang2025tunable}, and hot spot engineering~\cite{yang2023hot}.

The transfer of heat energy between bodies through the propagation of electromagnetic waves at nonzero temperatures is known as thermal radiation~\cite{howell2020thermal}. Over the past few decades, thermal radiation between two bodies, such as plates~\cite{ottens2011near,song2016radiative}, particles~\cite{narayanaswamy2008thermal,manjavacas2012radiative}, or disks~\cite{yu2017ultrafast,ramirez2017near,hu2020enhanced}, has gained considerable attention. This interest is driven by the potential for nanoscale heat flux to exceed the far-field limit predicted by Planck's black-body radiation law by several orders of magnitude, primarily owing to the additional contribution of evanescent waves~\cite{cuevas2018radiative}. The enhanced heat flux offers promising applications in waste heat recovery~\cite{jouhara2018waste,zhao2018near}, solar energy harvesting~\cite{ahmed2021review,fan2020near}, and nanoscale thermophotonic devices~\cite{ben2013phase,fiorino2018thermal,guo2020radiative}. Beyond the traditional focus on two-body systems, advances in extreme nanofabrication techniques have prompted numerous studies on thermal radiation in many-body systems~\cite{biehs2021near}. Recently, the integration of innovative concepts such as topology~\cite{liu2024topological}, nonreciprocity~\cite{zhu2018theory}, and twisted geometry~\cite{zheng2022molding} into ordered many-body systems has emerged as a viable strategy for achieving robust, directional, and tunable energy flow from a designated heat source. Despite existing discoveries have identified ballistic, diffusive, and anomalous diffusive transport regimes in many-body thermal radiation~\cite{ben2013heat,latella2018ballistic,tervo2019thermal,kathmann2018limitations}, the impact of disorder remains elusive due to challenges imposed by the intrinsic material loss, the vectorial nature of electromagnetic waves, and long-range many-body interactions within these subwavelength thermophotonic structures.

In our previous work~\cite{hu2024emergent}, we introduced a one-dimensional many-body chain exhibiting quasiperiodic order, which demonstrates Anderson localization of electric dipole modes at deep subwavelength scales. In this work, we demonstrate theoretically the suppression of radiative heat flux due to Anderson localization in such a quasiperiodic chain composed of plasmonic indium antimonide (InSb) nanoparticles. We first recall the eigenfrequency spectrum, the emergent localization transition, and the diverse localization phases within this chain. Subsequently, using many-body radiative heat transfer theory, we uncover that the presence of localized modes significantly reduces the transmission coefficient, thereby decreasing the radiative heat flux in the low-damping regime. We also provide direct evidence for suppression from Anderson localization by correlating the transmission coefficient with eigenmodes. Finally, we investigate how the suppression of radiative thermal conductance depends on interparticle spacing and damping rates.

\section{Results and discussion}

\begin{figure}[htbp]
\centering
\includegraphics[width=0.5\textwidth]{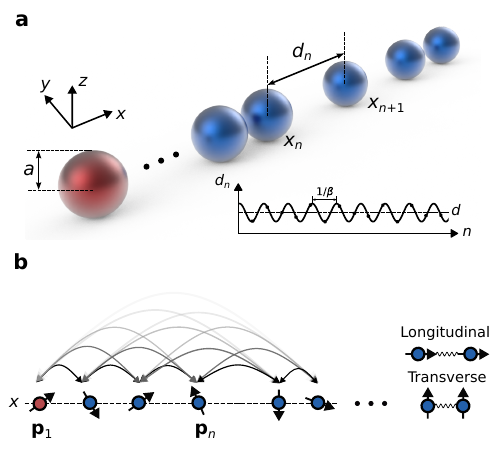}
\caption{\label{fig:1}\textbf{Sketch of one-dimensional quasiperiodic plasmonic dipole array.} \textbf{a} A finite chain of indium antimonide (InSb) nanoparticles with identical radii $a$, exhibiting Aubry-Andr\'{e}-Harper modulation in the interparticle spacing $d_n$. The inset illustrates the modulation profile and the distribution of spacings with black dots indicating the spacing between the $n^{\text{th}}$ and $n+1^{\text{th}}$ particles. \textbf{b} The chain is simplified as an array of dipoles with many-body interactions incorporated. The interaction strength is adjustable by modifying the nanoparticle spacings within the array. These interactions are decoupled into two distinct polarizations: longitudinal ($x$) and transverse ($y$ and $z$). In the thermal energy transport model, the first nanoparticle ( marked in red) is heated to a temperature of $T+\Delta T$, while the other nanoparticles (marked in blue) and the background are fixed at a constant temperature of $T$.}
\end{figure}

\textbf{Quasiperiodic plasmonic dipole chain.} The system under consideration is a one-dimensional chain of $N$ identical homogeneous, isotropic, and metallic spherical nanoparticles, which can be fabricated using advanced nanofabrication techniques~\cite{freire2025plasmonic}, such as electron-beam lithography, colloidal self-assembly, and nanoimprint lithography. These particles have radius $a$ and are aligned along the $x$-axis. As sketched in Fig.~\ref{fig:1}a, these nanoparticles are centered at positions $x_n$ and are separated by a center-to-center distance $d_n$, which obeys an artificial Aubry-Andr\'{e}-Harper (AAH) modulation~\cite{hu2024emergent}:
\begin{equation}
d_n=x_{n+1}-x_{n}=d\left[1+\eta\cos{(2\pi\beta n+\phi)}\right].
\end{equation}
Here, $x_n$ represents the position of the $n$th nanoparticle ($n$ is an integer and $1\le n\le N$) and $d$ is the average spacing between two adjacent particles. The parameter $\eta$ quantifies the strength of the cosine modulation, while $\beta$ governs its periodicity. If $\eta=0$, this is a periodic chain with lattice parameter $d$. For irrational $\beta$, the modulation is incommensurate with the lattice, corresponding to a quasiperiodic order. Whereas, for rational $\beta$, the modulation is commensurate, and the system can therefore be mapped onto a periodic lattice with a period of $q$ in the thermodynamic limit. This period is determined by the condition $q\beta\in \mathbb{Z}$, where $\mathbb{Z}$ is the set of integers. The phase factor $\phi$ can be regarded as the momentum in a synthetic orthogonal dimension, influencing the presence of topological edge modes~\cite{kraus2012topological,yan2025near}. Without loss of generality, we set $\beta=(\sqrt{5}-1)/2$ and $\phi=0$ in our numerical calculations.

The optical response of each metallic nanoparticle can be described by a Drude-type permittivity $\varepsilon$:
\begin{equation}    \varepsilon(\omega)=\varepsilon_{\infty}\left[1-\frac{\omega_{\text{p}}^{2}}{\omega(\omega+i\Gamma)}\right],
\end{equation}
where $\omega$ is the angular frequency, $\varepsilon_{\infty}$ is the high-frequency-limit permittivity, $\omega_{\text{p}}=\sqrt{n_\text{c} e^2/(m^{*}\varepsilon_0)}$ is the plasma frequency, $m^*$ is the doping-dependent effective mass of electrons, and $n_\text{c}$ is the doping concentration. Importantly, $\Gamma=e/(m^* \mu)$ is the Ohmic damping rate related to the absorptive loss of material, which can be controlled by the electron mobility $\mu$ in experiments~\cite{liu2021evolution}. Adopting a more realistic model including the contribution of phonon-polaritons for the permittivity does not affect our main conclusion~\cite{palik1976coupled,moncada2021near}. In the following, we adopt \textit{n}-type doped InSb as the constitutive material for the nanoparticles, with parameters~\cite{law2014doped}: $\varepsilon_{\infty}=15.68$, $n_\text{c}=1.36\times 10^{19}$ cm$^{-3}$, and $\omega_\text{p}=1.86\times 10^{14}$ rad s$^{-1}$. This model predicts a localized surface plasmon resonance at the spherical interface, occurring at the frequency $\omega_{\text{lsp}}=\omega_\text{p}\sqrt{\varepsilon_{\infty}/(\varepsilon_{\infty}+2)}\approx 1.7516\times 10^{14}$ rad s$^{-1}$, as per the Fr\"{o}hlich condition $\text{Re}[\varepsilon(\omega_{\text{lsp}})]=-2$. Consequently, each nanoparticle in the array supports three degenerate orthogonal localized surface plasmon resonances, polarized along the $\nu=x$, $y$, and $z$ directions, and characterized by the frequency $\omega_{\text{lsp}}$. The resonance clearly lies in the mid-infrared regime around $\lambda_{\text{lsp}}\approx 10.8$ $\upmu$m, which is relevant for thermal radiation around room temperature. In this work, we focus on the small-nanoparticle regime where the radius $a$ is smaller than $\lambda_{\text{lsp}}$, such that nanoparticles can be approximated as point dipoles. Their dipole polarizability, related to optical susceptibility in a vacuum, is given by,
\begin{equation}
    \alpha(\omega)=\frac{\alpha_{0}(\omega)}{1+i\gamma\alpha_{0}(\omega)},
\end{equation}
where $\alpha_0$ is the Lorenz-Lorentz polarizability of a nanoparticle with radius $a$, expressed in terms of permittivity $\varepsilon$ as,
\begin{equation}
    \alpha_0=4 \pi a^{3}\varepsilon_0\frac{\varepsilon-1}{\varepsilon+2},
\end{equation}
and $\gamma=-k^3/(6\pi\varepsilon_0)$ is the radiative correction term, which ensures energy conservation of the system by balancing scattering and extinction. $\varepsilon_0$ is the vacuum permittivity, $k=\omega/c$ is the vacuum wavenumber, and $c$ is the speed of light in vacuum.

We further assume that the interparticle distance $d_n$ exceeds $3a$, or equivalently, $\eta<1-3a/d$. This assumption enables us to safely neglect multipole effects and treat the nanoparticles as pure dipoles with dipole moments $\mathbf{p}_n$, as illustrated in Fig.~\ref{fig:1}b. Hence, the electromagnetic response of the InSb nanoparticle chain can be modeled using the self-consistent coupled dipole equations. And the eigenpairs of collective dipole eigenmodes can be determined
through a linearized Green’s function method. More details are listed in the “Coupled dipole equations” and “Eigenfrequencies and eigenvectors” subsections of Methods.

\begin{figure}[htbp]
\centering
 \includegraphics[width=\textwidth]{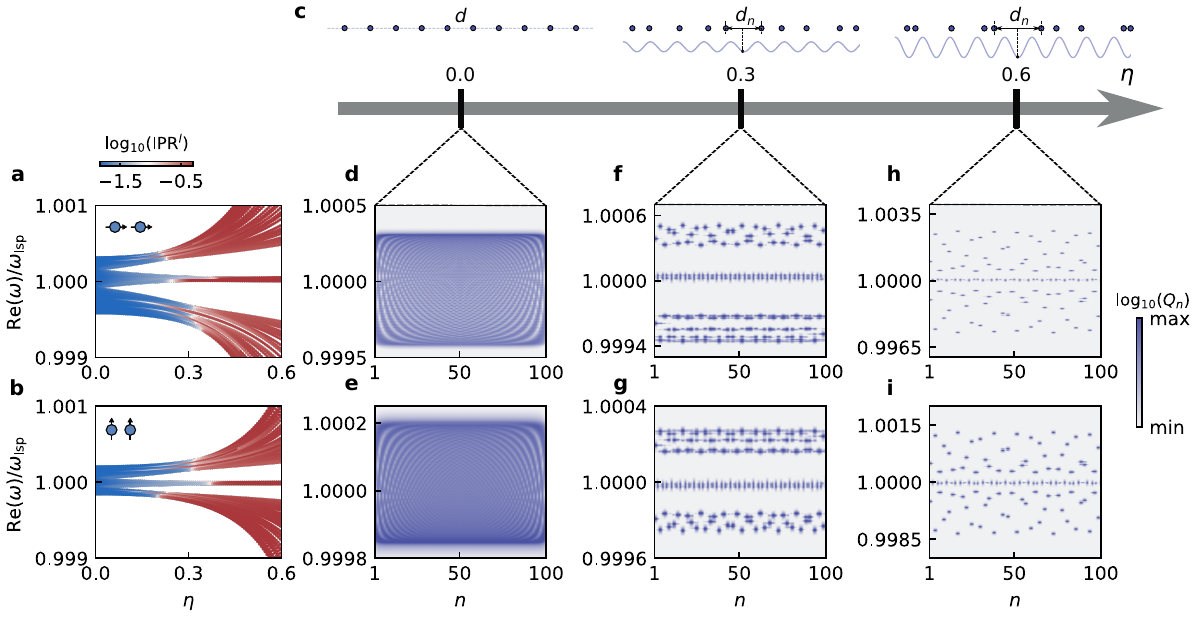}
\caption{\label{fig:2}\textbf{Emergent localized eigenmodes in quasiperiodic plasmonic arrays.} \textbf{a}, \textbf{b} Inverse participation ratio (IPR) of different eigenmodes as a function of corresponding eigenfrequencies and quasiperiodic modulation strength $\eta$ for longitudinal and transverse polarizations, respectively. \textbf{c} Schematic of array configurations at three modulation strengths: $\eta=0$, $\eta=0.3$, and $\eta=0.6$.
\textbf{d}-\textbf{i} Spatial distributions of effective extinction efficiency $Q_n$ for longitudinal (\textbf{d}, \textbf{f}, \textbf{h}) and transverse (\textbf{e}, \textbf{g}, \textbf{i}) polarizations at the corresponding modulation strengths. The parameters for the calculations are set as $N=100$, $a=20$ nm, $d=200$ nm, $\beta=(\sqrt{5}-1)/2$ and $\phi=0$. Absorptive loss can impede the observation of band modes through optical field excitation due to the skin effect of metallic nanoparticles. Thus, a small Ohmic damping of $\Gamma=1\times 10^{9}$ rad s$^{-1}$ is chosen to improve the resolution here.
}
\end{figure}

\textbf{Plasmonic dipole eigenmodes.}
We first demonstrate the presence of Anderson localization in these quasiperiodic nanoparticle chains. Figures~\ref{fig:2}a and \ref{fig:2}b display the eigenfrequency spectra for a chain with parameters $N=100$, $d=0.2$ $\upmu$m, and $a=20$ nm, considering both longitudinal and transverse polarizations with increasing levels of modulation strength $\eta$.

The first observation from Figs.~\ref{fig:2}a and \ref{fig:2}b is that, for both polarization types, the continuous spectrum splits into multiple subbands, separated by discernible gaps, as $\eta$ is turned on. This is a signature of the fractal spectrum, stemming from the irrational nature of the AAH modulation. In the thermodynamic limit, this spectrum forms a Cantor set, which can be constructed by iteratively dividing intervals\cite{DasSarma_PRL2010}. Furthermore, the spectrum is centered around the resonant frequency of single nanoparticle $\omega_{\text{lsp}}$, and shifts to both higher and lower frequency regions as $\eta$ increases, thereby exhibiting a larger bandwidth. However, the spectrum is asymmetric with respect to $\omega_{\text{lsp}}$ due to the breaking of chiral symmetry by long-range many-body interactions. Specifically, in Fig.~\ref{fig:2}a, for the longitudinal polarization, the low-frequency subbands have a wider bandwidth compared to the high-frequency subbands, whereas the opposite is true for the transverse polarization shown in Fig.~\ref{fig:2}b. This feature can determine the spectral transmission window location of dipole gases under the condition of a large Ohmic damping rate, as will be discussed later.

From the color coding  representing the inverse participation ratio (IPR, see the “Inverse participation ratio” subsection of Methods) of the corresponding eigenvectors, we can also see that, as $\eta$ increases, localized dipole modes (red, high IPR) emerge when $\eta$ exceeds a critical value $\eta_\text{c}$, clearly separated from the extended modes (blue, low IPR). Note that for the longitudinal polarization, localized modes arise from the high-frequency band edge, while for the transverse polarization, they debut at the low-frequency band edge. These localized modes induced by Anderson localization are expected to suppress the radiative heat flux in quasiperiodic configurations compared to their periodic counterparts. We also identify specific localized modes within the fractal gaps. These modes, emerging in the presence of minor disorder, localize at the edges and possess topological protection by topological invariants.

To substantiate the localization behavior observed in the diagram of the eigenfrequency spectrum, we calculate the effective extinction efficiency $Q_n$ (see the “Effective extinction efficiency” subsection of Methods) for each nanoparticle in the arrays~\cite{zundel2022green}.

We summarize the effective extinction efficiency in Figs.~\ref{fig:2}d-i as a function of frequency $\omega$ and site $n$ for three different array configurations. In the disorder-free limit ($\eta=0$), corresponding to a periodic array, shown in Figs.~\ref{fig:2}d and \ref{fig:2}e, the extinction efficiency $Q_n$ for both polarizations displays typical Bloch wave patterns, asymmetrically distributed around $\omega_{\text{lsp}}$ within the frequency window, akin to the eigenfrequency spectra. By further increasing $\eta$ to a moderate level, e.g., $\eta=0.3$, the extinction efficiency shows a disconnected, hyphen-like pattern in the high-frequency region for longitudinal polarization (Fig.~\ref{fig:2}f) and in the low-frequency region for transverse polarization (Fig.~\ref{fig:2}g). This observation suggests the presence of localized dimer modes, where localized surface plasmon resonances are excited in two adjacent nanoparticles. In this context, the disordered dipole array demonstrates an intermediate phase where localized and extended modes coexist. With strong disorder ($\eta=0.6$), the array transitions into a fully localized phase, as evidenced by the hyphen-like pattern spanning the entire frequency range for both polarizations (Fig.~\ref{fig:2}h-i).

\begin{figure}[htbp]
\centering
 \includegraphics[width=\textwidth]{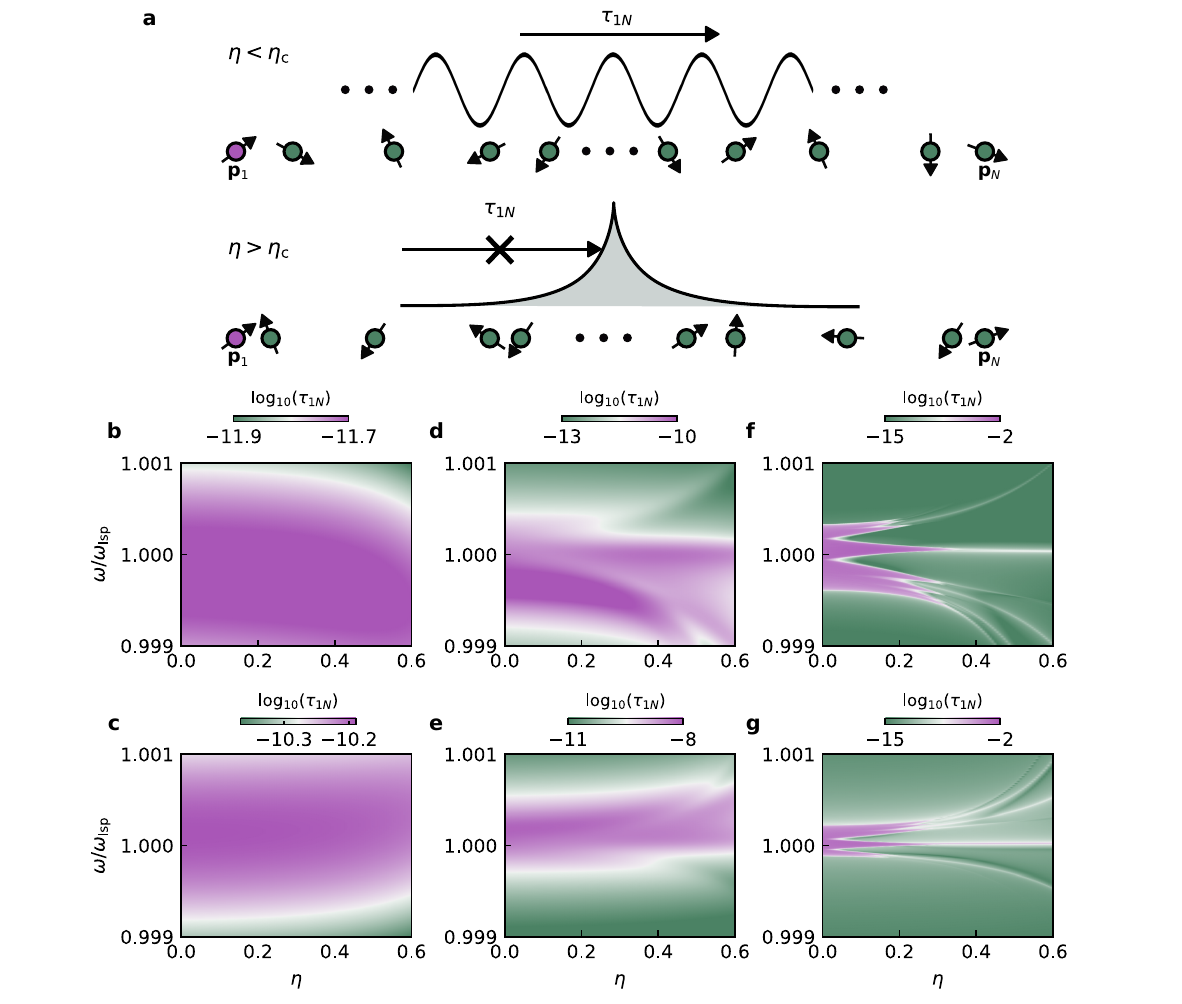}
\caption{\label{fig:3}\textbf{Transmission spectra of quasiperiodic plasmonic arrays.} \textbf{a} Illustration of transmission channels between the first and final dipoles for a specific frequency in relation to the Anderson localization transition. Before the localization transition ($\eta<\eta_\text{c}$), plasmonic dipoles form an extended collective wave as sketched by the black curve, while after the localization transition ($\eta>\eta_\text{c}$), a localized collective wave emerges, as denoted by the gray-shaded curve. \textbf{b}-\textbf{g} Spectral transmission coefficient $\tau_{1N}$ between dipoles at the chain's two ends as a function of modulation strength $\eta$ for  longitudinal (\textbf{b}, \textbf{d}, \textbf{f}) and transverse (\textbf{c}, \textbf{e}, \textbf{g}) polarizations with different Ohmic damping rates: (\textbf{b}, \textbf{c}) $\Gamma=1\times 10^{12}$ rad s$^{-1}$, (\textbf{d}, \textbf{e}) $\Gamma=5\times 10^{10}$ rad s$^{-1}$, and (\textbf{f}, \textbf{g}) $\Gamma=1\times 10^{9}$ rad s$^{-1}$. Other parameters are the same with those in Fig.~\ref{fig:2}.}
\end{figure}

\textbf{Spectral transmission coefficient.} We now proceed to illuminate the impact of Anderson localization on radiative heat transfer in the quasiperiodic array. The initial nanoparticle at the left end of the chain functions as a dipole emitter, while the final nanoparticle at the right end serves as the receiver. A crucial ingredient of our transport analysis is the transmission coefficient $\tau_{1N}(\omega)$ from nanoparticle 1 to $N$ (The general formalism is derived in Supplementary Note 1, which includes Fig.~S1\cite{novotny2012principles,messina2013fluctuation,cxb_small2018,cxb_PRB2019,yan2025,ashida2020non,ott2021thermal,yang2025non}).

For a fixed emission frequency, one can anticipate that the energy transmission between the emitter and the receiver should be supported if the system exhibits an extended collective mode. Once the modulation strength $\eta$ exceeds a critical value ($\eta_\text{c}$) such that the system undergoes the localization transition due to the destructive interference of collective modes,
localized collective modes should emerge and the transmission channel closes subsequently, as illustrated in Fig.~\ref{fig:3}a.

To quantitatively assess the impact of Anderson localization, we compute the spectral transmission coefficient $\tau_{1N}$ as the modulation strength $\eta$ increases for a quasiperiodic array of $100$ nanoparticles at three distinct Ohmic damping rates: $\Gamma=1\times 10^{12}$, $5\times 10^{10}$, and $1\times 10^{9}$ rad s$^{-1}$, as shown in Figs.~\ref{fig:3}b-\ref{fig:3}g. The geometrical parameters are identical to those used in the calculation of collective eigenmodes in Figs.~\ref{fig:2}a and \ref{fig:2}b. In the case of natural damping rate ($\Gamma=1\times 10^{12}$ rad s$^{-1}$), as depicted in Figs.~\ref{fig:3}b and \ref{fig:3}c, the transmission of longitudinal modes is primarily contributed by band modes in the low-frequency region (Fig.~\ref{fig:3}b), whereas the transmission of transverse modes is predominantly attributed to the high-frequency band modes (Fig.~\ref{fig:3}c). The asymmetric transmission spectra are consistent with the asymmetric profiles of the eigenfrequency spectra and local effective extinction spectra for both polarizations, as discussed in Fig.~\ref{fig:2}. However, due to the large intrinsic damping rate in InSb nanoparticles, the band modes are significantly broadened, complicating the identification of contributions from individual band modes. Consequently, the role of localized modes is obscured, even in systems with extensive quenched disorder.

By reducing the damping rate to $\Gamma=5\times 10^{10}$ rad s$^{-1}$, the contribution of different band modes to the transmission spectrum allows for a better discrimination as shown in Figs.~\ref{fig:3}d and \ref{fig:3}e. If we further decrease the damping rate by a factor of $50$, i.e., $\Gamma=1\times 10^{9}$ rad s$^{-1}$, the responses of band modes no longer overlap. This yields a discernible transmission spectrum in Figs.~\ref{fig:3}f and \ref{fig:3}g that aligns with the eigenfrequency spectrum. The suppression effect of localized modes on transmission becomes evident for both polarizations when the plasmonic array enters into the localized phase. The contributions of topological edge modes can also be identified from the white trajectories within the transmission gaps, which display slightly higher values than those of localized bulk modes. This is due to the fact that these topological edge modes are localized at the two ends of the chain.
\begin{figure}[htbp]
\centering
 \includegraphics[width=\textwidth]{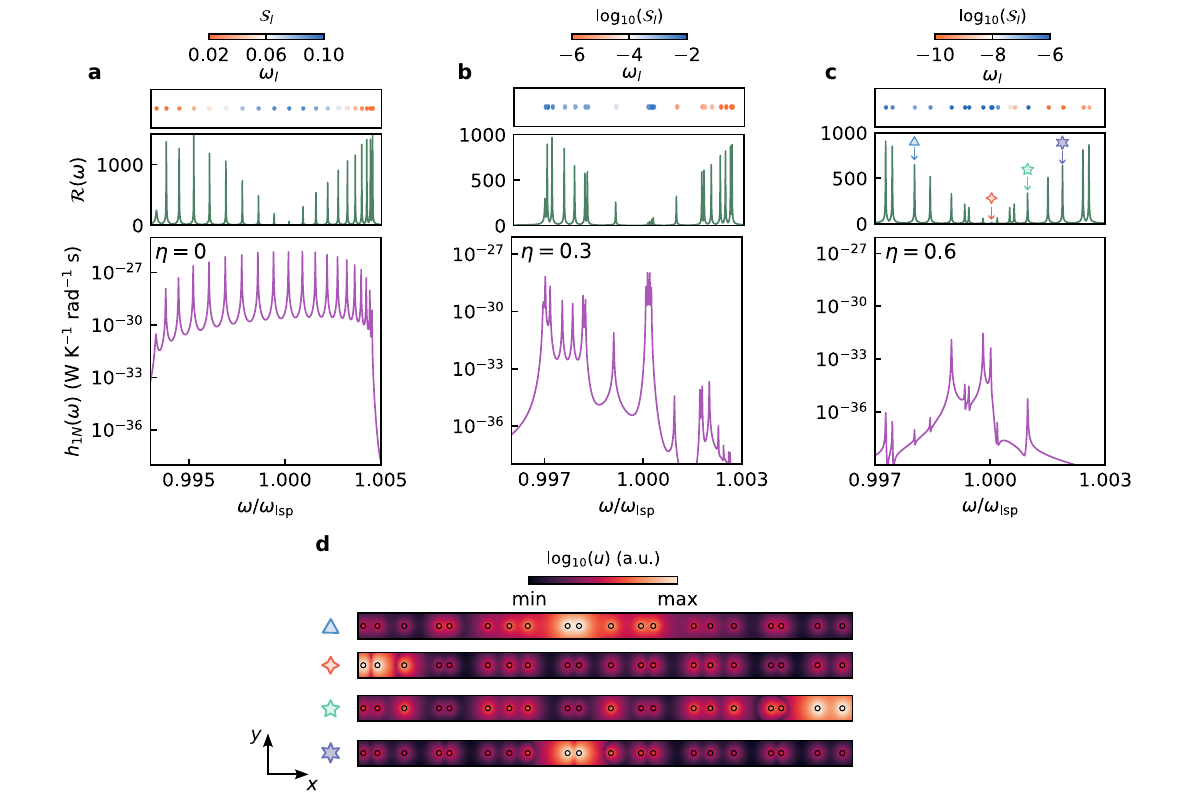}
\caption{\label{fig:4}\textbf{Radiative heat transport through eigenchannels.} \textbf{a}-\textbf{c} Eigenfrequency spectrum $\omega_l$ colored with localization index $\mathcal{S}_l$, spectral response function $\mathcal{R}(\omega)$, and spectral thermal conductance $h_{1N}(\omega)$ between nanoparticles at the chain's two ends for the longitudinal polarization at three given modulation strengths of $\eta=0$, $\eta=0.3$, and $\eta=0.6$, respectively. The radii of the nanoparticles are specified as \textbf{a} $a=50$ nm, \textbf{b} $35$ nm, and \textbf{c} $20$ nm. The star markers denote the cases of excited frequencies discussed in \textbf{d}. \textbf{d} Spatial contours of the radiated electric-field energy density $u$, evaluated at a height of 30 nm above the array for the longitudinal polarization, corresponding to the cases in \textbf{c}. The other parameters include $N=20$, $d=200$ nm, $T=300$ K, $\beta=(\sqrt{5}-1)/2$, and $\phi=0$.}
\end{figure}

\textbf{Mode analysis of spectral heat flux.} To better understand the impact of Anderson localization on radiative energy transfer, we establish a connection between eigenmodes and transmission channels. The transmission coefficient $\tau_{1N}$ can be decomposed into the contributions of all eigenmodes for a single polarization in the following fashion (see Supplementary Note 2 for details):
\begin{equation}
\tau_{1N}= 4\frac{\chi^2}{|\alpha|^2}\left|\sum_{l=1}^{N} \frac{p^l_{N}p^l_{1}}{\sum_{n'=1}^{N}\left(p_{n'}^l\right)^2}\frac{1}{1-\alpha\lambda_l}\right|^2  =4\frac{\chi^2}{|\alpha|^2}\left|\sum_{l=1}^{N} S_l R_l\right|^2,
\end{equation}
where $S_{l}$ is the spatial variation of dipole moment component for the $l$th eigenmode, and $R_l$ is the response frequency variation of the eigenmode. We define a spectral response function $\mathcal{R}(\omega)$ as
\begin{equation}
\mathcal{R}(\omega)=\max_{l}\left|R_l(\omega)\right| =\max_{l}\left|\frac{1}{1-\alpha(\omega)\lambda_l(\omega)} \right|,
\end{equation}
and introduce a spatial localization index $\mathcal{S}_l$ for nanoparticles positioned at two ends:
\begin{equation}
\mathcal{S}_l=\left|S_{l} \right|=\left|\frac{p^l_{N}p^l_{1}}{\sum_{n'=1}^{N}\left(p_{n'}^l\right)^2}\right|.
\end{equation}

Figures~\ref{fig:4}a-c display the spectral response function $\mathcal{R}(\omega)$ and the localization index $\mathcal{S}_{l}$ corresponding to the eigenfrequencies $\omega_l$ for longitudinal modes of a chain with $d=0.2$ $\upmu$m and $N=20$. These are plotted for three different modulation strengths: $\eta=0$, $0.3$, and $0.6$, respectively. Additionally, we present the spectral thermal conductance $h_{1N}(\omega)$ between the chain's two ends for the longitudinal polarization at a temperature of $T=300$ K, within the same frequency window. The radiative thermal conductance $h_{1N}(\omega)$ for a given frequency $\omega$ can be expressed via the transmission coefficient $\tau_{1N}$:
\begin{equation}
 h_{1N}(\omega)=\frac{\hbar \omega}{2 \pi}\frac{\partial f_{\text{BE}}(\omega, T) }{\partial T} \tau_{1N}(\omega),
\end{equation}
where $f_{\text{BE}}(\omega, T)=1/[e^{\hbar\omega/(k_{\text{B}}T)}-1]$ is Bose-Einstein distribution at frequency $\omega$ and temperature $T$. In the valid interval of the dipole approximation, the radii of the nanoparticles are configured as $a=50$ nm, $35$ nm, and $20$ nm to achieve the broadest possible spectrum. To reveal the link between eigenmodes and heat flux spectrum unambiguously, we probe into the low-damping limit of $1\times 10^{9}$ rad s$^{-1}$, where all resonant channels are well-resolved.

In Figs.~\ref{fig:4}a-c, the spectral response function presents a clear structure composed of non-overlapping Lorentzian curves, with resonant peaks matching well with the eigenfrequencies of the normal modes for all levels of disorder. The resonant peaks precisely corresponding to the eigenfrequencies are also observed in the spectral thermal conductance. In the ordered case of $\eta=0$, the spectral conductance is negligible for most frequencies, except those near the eigenfrequencies. This is not the case for $\eta=0.3$, where the presence of localized modes above the mobility edge $\omega_\text{c}\approx 1.001\omega_{\text{lsp}}$ diminishes the spectral conductance. These localized modes manifest small localization indices, which assign low weights to the Lorentzian curves. For $\eta=0.6$, the system is filled with localized modes, and thus, the spectral conductance is drastically suppressed across the entire frequency window, particularly in the high-frequency region. A small peak manifests at a frequency of $1.001\omega_{\text{lsp}}$. This frequency corresponds to a topological edge mode, which shows a slightly higher localization index than the localized bulk modes. We also analyzed transverse modes, which exhibit analogous characteristics to longitudinal modes, albeit with a narrower and inverted spectrum.

\begin{figure}[htbp]
\centering
 \includegraphics[width=\textwidth]{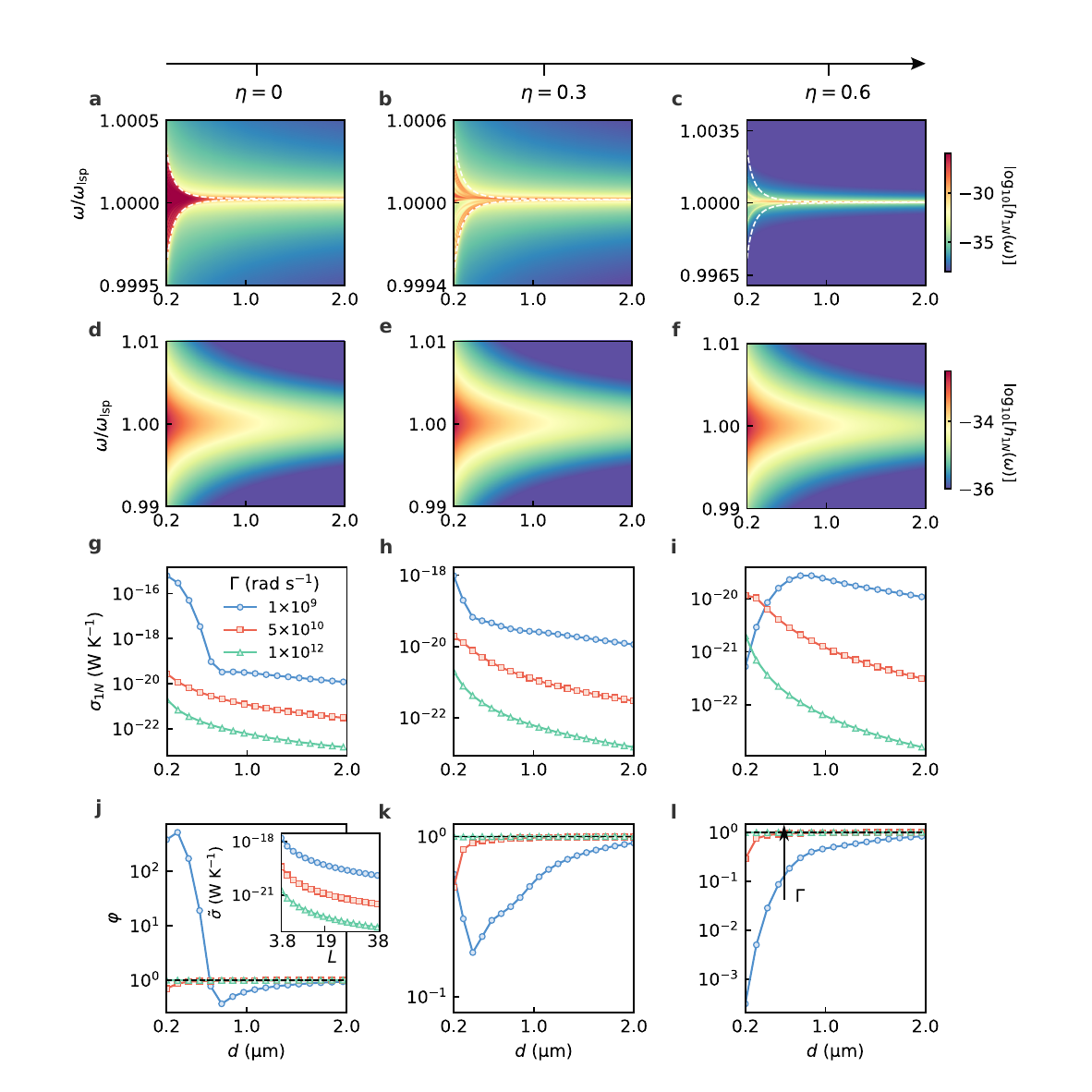}
\caption{\label{fig:5}\textbf{Radiative heat transfer mediated by many-body interactions.} \textbf{a}-\textbf{f} Spectral thermal conductance $h_{1N}(\omega)$ for the integrated contributions of longitudinal and transverse polarizations as a function of the mean interparticle distance $d$ at fixed modulation strengths of $\eta=0$ (\textbf{a}, \textbf{d}), $\eta=0.3$ (\textbf{b}, \textbf{e}), and $\eta=0.6$ (\textbf{c}, \textbf{f}) with damping rates of \textbf{a}-\textbf{c} $1\times 10^{9}$ rad s$^{-1}$ and \textbf{d}-\textbf{f} $1\times 10^{12}$ rad s$^{-1}$. \textbf{g}-\textbf{l} Total thermal conductance $\sigma_{1N}$ and modulation ratio $\varphi$ versus interparticle distance $d$, evaluated for three modulation strengths and damping rates. The inset in (\textbf{j}) gives the baseline case of two-body thermal conductance. The black dashed line indicates $\varphi=1$. All calculations assume $N=20$, $a=20$ nm, $d=200$ nm, $T=300$ K, $\beta=(\sqrt{5}-1)/2$, and $\phi=0$.}
\end{figure}

In order to clarify the modes responsible for the resonant peaks, we employ the local density of electric field energy to illustrate the spatial confinement of energy emitted by dipoles at resonant frequencies of interest. The density of radiated electric field energy, denoted as $u(\mathbf{r}, \omega)$, at a field point $\mathbf{r}$ above the chain at temperature $T$ can be calculated according to Equation~(35) in Supplementary Note 3. Figure~\ref{fig:4}d showcases the spatial distribution of energy density on the plane at $z=30$ nm for longitudinal modes at a strong disorder level of $\eta=0.6$ and a temperature of $T=300$ K. The four panels correspond to the four excited frequencies: $\omega\approx 0.998\omega_{\text{lsp}}$, $\omega_{\text{lsp}}$, $1.001\omega_{\text{lsp}}$, and $1.002\omega_{\text{lsp}}$, as marked by the triangle and star symbols in Fig.~\ref{fig:4}c. The Ohmic damping rate is maintained at $\Gamma=1\times 10^{9}$ rad s$^{-1}$ for visibility. For the localized bulk modes at $0.998\omega_{\text{lsp}}$ and $1.002\omega_{\text{lsp}}$, the energy is confined at the two middle nanoparticles in the chain, suggesting dimer localization. For the other two frequencies, $\omega_{\text{lsp}}$ and $1.001\omega_{\text{lsp}}$, the energy distribution is concentrated at the two ends, indicating topological edge localization. In a nutshell, these results demonstrate that the resonant peaks are governed by localized bulk and edge modes in the presence of strong disorder. Further analysis of larger damping rates is presented in Supplementary Note 4 and Figs.~S2-S3, which underscore spectral overlap between modes in the response function, conductance, and local energy density.

\textbf{Suppression in thermal conductance.} Figures~\ref{fig:5}a-f illustrate the dependence of the spectral thermal conductance $h_{1N}(\omega)$ on the mean interparticle spacing $d$, incorporating all polarization components, under low ($\Gamma=1\times 10^9$ rad s$^{-1}$) and high ($\Gamma=1\times 10^{12}$ rad s$^{-1}$) Ohmic damping rates. Results are shown for nanoparticle arrays with $N=20$ and $a=20$ nm at three modulation strengths: $\eta=0$, $\eta=0.3$, and $\eta=0.6$. As observed in Fig.~\ref{fig:5}a-c, for the low damping rate ($\Gamma=1\times 10^9$ rad s$^{-1}$), the thermal conductance spectra are bounded by the maximum and minimum eigenfrequency bands of these arrays, as indicated by the white dashed curves. As $d$ increases to 0.6 $\upmu$m, all bulk bands merge at the localized surface plasmon frequency $\omega_{\text{lsp}}$, which boils down to the decoupling of plasmonic dipoles. In essence, with greater interparticle spacing, the many-body interactions between dipoles become weaker, leading to diminished collective behavior. Ultimately, in the large spacing regime $d>0.6$ $\upmu$m, the bands get very narrow and the eigenfrequencies tend to become degenerate at $\omega_{\text{lsp}}$. Hence, for all $\eta$ cases, the spectral conductance only peaks at $\omega_{\text{lsp}}$ in this regime, as depicted in Figs.~\ref{fig:5}a-c. Whilst, in the small-spacing regime ($d< 0.6$ $\upmu$m), many-body effects are pronounced in the spectral conductance for all the configurations, particularly evident in the enhancement at $\eta=0$ and the suppression at $\eta=0.6$, as shown in Figs.~\ref{fig:5}a and \ref{fig:5}c, respectively. However, in the overdamping scenario ($\Gamma=1\times 10^{12}$ rad s$^{-1}$) shown in Fig.~\ref{fig:5}d-f, all $\eta$ cases display identical distributions, with many-body effects becoming unobservable.

We now analyze the variation of total thermal conductance $\sigma_{1N}$ with respect to the mean spacing $d$ for the above arrays. $\sigma_{1N}$ is obtained by integrating contributions across all frequencies and summing over all polarizations. It is anticipated that significant modulation, due to many-body interactions, will manifest in the small-spacing, low-damping region. Figures~\ref{fig:5}g-i present the results calculated for three distinct damping rates $\Gamma$, ranging from $1\times 10^{9}$ to $1\times 10^{12}$ rad s$^{-1}$. At $\Gamma=1\times 10^{9}$ rad s$^{-1}$, the total thermal conductance drops sharply from $\eta=0$ to $\eta=0.6$ in the small-spacing regime, whereas differences in conductance narrow at larger $d$. The strong disordered case ($\eta=0.6$) demonstrates a local maximum in thermal conductance at intermediate $d$, attributed to the Anderson localization at small $d$. The situation is different, however, when the nanoparticle in the array dissipates strongly. Conductance disparities diminish with increasing $\Gamma$ for all array configurations, as overdamping erases many-body effects.

\begin{figure}[htbp]
\centering
 \includegraphics[width=\textwidth]{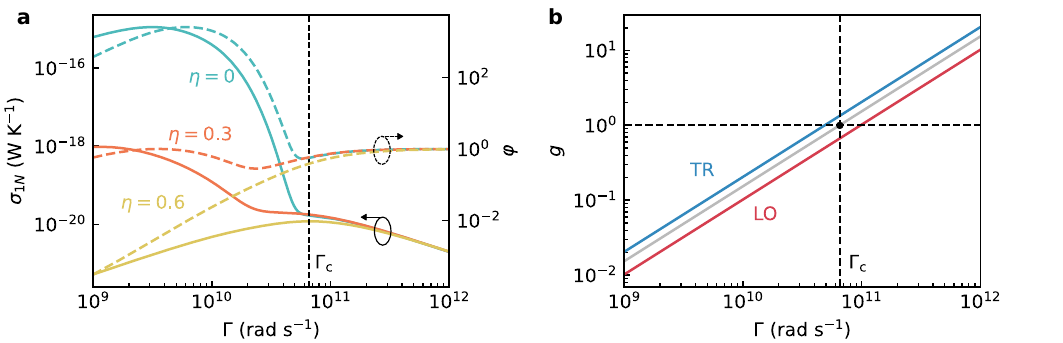}
\caption{\label{fig:6}\textbf{Effect of Ohmic damping rates on thermal conductance.} \textbf{a} Total thermal conductance $\sigma_{1N}$ (solid curves) and modulation ratio $\varphi$ (dashed curves) as a function of Ohmic damping rate $\Gamma$ for different values of modulation strength $\eta$. \textbf{b} Thouless number $g$ as a function of $\Gamma$ at $\eta=0.6$ for longitudinal polarization (LO, red line), transverse polarization (TR, blue line), and the average of both polarizations (gray line). The dashed vertical and horizontal lines indicate the localization thresholds at $\Gamma_\text{c}=6.5\times 10^{10}$ rad s$^{-1}$ and $g_\text{c}=1$, respectively. The other parameters for both plots. The parameters for both plots are set as follows: $N=20$, $a=20$ nm, $d=200$ nm, $T=300$ K, $\beta=(\sqrt{5}-1)/2$, and $\phi=0$.}
\end{figure}

To quantify the modulation in total thermal conductance caused by many-body interactions, we define a modulation ratio $\varphi$ as:
\begin{equation}
\varphi=\frac{\sigma_{1N}}{\tilde{\sigma}},
\end{equation}
where $\tilde{\sigma}$ represents the total radiative thermal conductance of two isolated nanoparticles positioned at the opposing ends of the chain, computed with two-body interactions only. Note that the separation distance between the two nanoparticles fluctuates with varying levels of modulation strength. Nonetheless, the thermal conductance remains almost the same across these distances. Therefore, the thermal conductance in an average separation distance of $L=(N-1)d$ is deemed as the benchmark, which can be approximated as~\cite{asheichyk2022radiative}
\begin{equation}
\tilde{\sigma}(L) \approx \frac{1}{4\pi^3}\int_{0}^{+\infty}\hbar\omega\frac{\partial f_{\text{BE}}(\omega, T)}{\partial T}k^4\chi^2 \left( \frac{1}{L^2} + \frac{1}{k^2 L^4} + \frac{3}{k^4 L^6} \right),
\end{equation}
where $\chi=\text{Im}(\alpha)+\gamma\left|\alpha\right|^{2}$ is the susceptibility of a single nanoparticle.
This baseline case is illustrated in the inset of Fig.~\ref{fig:5}j. It is observed that the conductance scales as $L^{-2}$ when the separation distance $L$ exceeds $L>\lambda_{\text{lsp}}$, reflecting far-field heat exchange. Figures~\ref{fig:5}j-l depict the modulation ratio $\varphi$ as a function of spacing $d$ for three distinct damping rates $\Gamma$. Each panel corresponds to a different value of $\eta$: $\eta=0$ (Fig.~\ref{fig:5}j), $\eta=0.3$ (Fig.~\ref{fig:5}k), and $\eta=0.6$ (Fig.~\ref{fig:5}l). In the case of $\Gamma=1\times 10^{9}$ rad s$^{-1}$, upon examining these results, we observe that, for $\eta=0$, when the array is in the extended phase, $\varphi$ shows a two-orders-of-magnitude enhancement, persisting up to a distance of $d\approx 0.6$ $\upmu$m ($L\approx \lambda_{\text{lsp}}$). Moreover, for $\eta=0.3$, in the intermediate phase, $\varphi$ shows a slight suppression rate. A pronounced suppression of three orders of magnitude manifests for $\eta=0.6$, leading to ultralow radiative heat transfer and indicating a transition to the localized phase. Importantly, as $\Gamma$ increases to $1\times 10^{12}$ rad s$^{-1}$, both enhancement and suppression effects shrink rapidly.

The above results clearly indicate that the damping rate is a limiting factor in the modulation of radiative heat flux mediated by many-body effects. This impact is further elucidated by plotting the total thermal conductance $\sigma_{1N}$ and modulation ratio $\varphi$ as functions of damping rate $\Gamma$ in Fig.~\ref{fig:6}a.
As we can see, a decrease in the damping rate significantly enhances or suppresses the thermal conductance, especially when the damping rate is below a threshold value ($\Gamma_\text{c}\approx 6.5\times 10^{10}$ rad s$^{-1}$). This occurs because decreasing the damping rate improves the mode quality of collective plasmon resonances. Consequently, a higher damping rate, which broadens the lineshape of lattice resonances, obstructs the sensitivity of radiative thermal conductance to mode shape changes, or more specifically, to variations in geometrical arrangements. This is reflected in the slight change in $\varphi$ when most of the modes overlap in the spectrum, i.e., $\Gamma>\Gamma_\text{c}$. To measure the degree of spectral mode overlap in the localized phase, we introduce the Thouless number $g$, defined as~\cite{wang2011transport,skipetrov2014absence}
\begin{equation}
g=\frac{\langle\delta\omega\rangle}{\langle\Delta\omega\rangle},
\end{equation}
where $\langle\delta\omega\rangle$ and $\langle\Delta\omega\rangle$ are the average decay rate of eigenmodes and average level spacing between adjacent modes, given by,
\begin{equation}
\langle\delta\omega\rangle=\frac{1}{N}\sum_{l}\text{Im}(\omega_l), \ \ \langle\Delta\omega\rangle=\frac{\text{Re}(\omega_N)-\text{Re}(\omega_1)}{N-1}.
\end{equation}
Here, the eigenmodes are sorted by the real part of eigenfrequencies in ascending order. The Thouless number $g$ is a fundamental localization parameter, with the diffusive and the localized regime being hallmarked by overlapping modes ($g>1$) and well-resolved modes ($g<1$), respectively. We plot the Thouless number $g$ as a function of damping rate $\Gamma$ at $\eta=0.6$ for longitudinal polarization (red), transverse polarization (blue), and the average of both polarizations (gray) in Fig.~\ref{fig:6}b. When $g<1$, corresponding to $\Gamma<\Gamma_\text{c}$,  different localized modes are distinguishable, which concurs with the observed decrease in thermal conductance in Fig.~\ref{fig:6}a. In this scenario, typically at a given frequency, only a single or a few localized modes participate in the transport process. The thermal conductance is consequently suppressed due to resonant tunneling through localized modes. Generally, the Ohmic damping rate is determined by electron scattering,  with typical values for realistic plasmonic materials on the order of $\Gamma\sim 10^{-3}\omega_{\text{lsp}}$. This rate may intensify if the array is supported by a substrate. Therefore, nanoparticles in experimental setups are expected to be more lossy, resulting in lower mode quality of collective resonance. Future work should focus on exploring plasmonic materials with reduced Ohmic losses, which will be crucial for achieving an observable suppression of radiative heat transfer caused by Anderson localization. In practical applications, the utilization of gain media is an alternative strategy for offsetting Ohmic losses~\cite{freire2025plasmonic}. Although we focused on chains composed of metallic nanoparticles in this work, it is worth noting that our model can also be extended to polar materials that support phonon-polaritons (See Supplementary Note 5 and Table~S1 for details).

\section{Conclusions}
In summary, we investigated the radiative heat transfer in a one-dimensional plasmonic chain with AAH-type spacing disorder. We demonstrated that an Anderson localization transition occurs in this chain, leading to localized plasmon modes that can significantly suppress radiative thermal conductance under conditions of low Ohmic dissipation. We validated the emergence of localization transition by analyzing the spatial profile of eigenmodes through the inverse participation ratio and local extinction efficiency for different values of modulation strength $\eta$ and damping rate $\Gamma$. We also linked spectral thermal conductance with plasmonic eigenmodes using mode expansion of the transmission coefficient to figure out the contribution of different eigenmodes to thermal conductance. Furthermore, we studied the dependence of thermal conductance on spacing, which exhibits a local maximum before asymptotically approaching the two-body limit in the Anderson localized phase. The suppression of thermal conductance is sensitive to the Ohmic damping rate, occurring only when the localized modes are well-resolved, characterized by a Thouless number less than one ($g<1$). Finally, we found that a smaller damping rate enhances the suppression of heat flux, with suppression rates exceeding three orders of magnitude at a damping rate of $1\times 10^9$ rad s$^{-1}$. Our work elucidates the impact of Anderson localization on radiative heat transfer and provides a fundamental understanding of many-body interactions in thermoplasmonics.

\section{Methods}
\subsection{Coupled dipole equations}
  The induced dipole $\mathbf{p}_n$ at the $n$th nanoparticle due to all other dipoles can be expressed as
\begin{equation}
 \mathbf{p}_n=\frac{k^{2}}{\varepsilon_{0}} \alpha(\omega) \sum_{n^{\prime} \neq n} \mathbf{G}\left(\mathbf{r}_n-\mathbf{r}_{n'}, \omega \right) \mathbf{p}_{n^{\prime}},
\end{equation}
where $n$ and $n'$ run over all $N$ nanoparticles, $\mathbf{r}_n$ is the position vector of $n$th dipole, $\mathbf{p}_{n}$ is the excited dipole moment of $n$th dipole, and $\mathbf{G}$ is the free-space dyadic Green's function describing the interactions between dipoles, which is given by~\cite{novotny2012principles}
\begin{equation}\label{eq:dyadic}
\mathbf{G}(\mathbf{r}, \omega) = \frac{\exp(ikr)}{4\pi r} \left[ \left( \frac{i}{kr} - \frac{1}{k^2r^2} + 1 \right) \mathbf{I}  + \left( -\frac{3i}{kr} + \frac{3}{k^2r^2} - 1 \right) \hat{\mathbf{r}} \otimes \hat{\mathbf{r}} \right],
\end{equation}
with $\mathbf{I}$ being an identity tensor, $r=|\mathbf{r}|>0$ being the distance between two arbitrary field points, and $\hat{\mathbf{r}}=\mathbf{r}/r$ being the unit vector parallel to $\mathbf{r}$. The free-space dyadic Green's function for the one-dimensional quasiperiodic array can be decoupled into the longitudinal ($x$) and transverse ($y$ or $z$) polarization components, denoted as $G_{xx}$ and $G_{yy/zz}$, which are given by
\begin{equation}\label{gf-xx}
G_{xx}(x, \omega)=\frac{2 e^{i kx}}{4 \pi k^{2} x^{3}}(1-i k x),
\end{equation}
\begin{equation}\label{gf-yz}
G_{y y / z z}(x, \omega)=\frac{-e^{i k x}}{4 \pi k^{2} x^{3}}\left(1-i k x-k^{2} x^{2}\right).
\end{equation}
The polarization-dependent Green's functions describe the anisotropic interactions between dipole pairs. In our coupled dipole model, the many-body interactions between different dipoles, included in Green's functions, can result in the coupling of individual dipoles into collective dipoles that extend over the entire chain. The eigenpairs of collective dipole eigenmodes can be determined through a linearized Green's function method.

\subsection{Eigenfrequencies and eigenvectors}
For a one-dimensional linear chain of $N$ dipoles, according to the polarization direction of the dipole moments, the eigenmodes can be divided into two types: transverse and longitudinal. We construct an interaction matrix $\mathbf{M}$ to reformulate the coupled dipole equations into a corresponding eigenvalue equation for longitudinal or transverse polarization,
\begin{equation}\label{eigeneq}
\mathbf{M}\mathbf{p}=\alpha^{-1}\mathbf{p},
\end{equation}
where $\mathbf{M}$ is the $N\times N$ interaction matrix, which contains all the information of dipole-dipole interactions between any two nanoparticles. Here we should keep in mind that the entries of interaction matrix $\mathbf{M}$ are filled with free-space Green's functions, corresponding to either longitudinal or transverse polarization, as defined in Eq.~(\ref{gf-xx}) or Eq.~(\ref{gf-yz}). For example, in the case of longitudinal polarization, the off-diagonal elements are given by $M_{nm}=\varepsilon_0^{-1}k^2 G_{xx}(x_{nm})$ with $x_{nm}=|x_n-x_m|$ being the distance between dipole $n$ and $m$ for $n\neq m$, while the diagonal elements are $M_{nn}=0$. The right eigenvector $\mathbf{p}=[p_{1}, p_{2}, \dots, p_{N}]^{\mathrm{T}}$ stands for the dipole moment distribution of an eigenmode. The eigenfrequencies are dictated by identifying the roots of the following equation,
\begin{equation}\label{secular-eq}
\det\left[\alpha^{-1}(\omega)\mathbf{I}-\mathbf{M}(\omega)\right]=0.
\end{equation}
Once an eigenfrequency $\omega_l$ is obtained, it is substituted into Eq.~(\ref{eigeneq}), which yields the corresponding eigenvector $\mathbf{p}_{l}$ as a nontrivial solution of this equation. The roots of Eq.~(\ref{secular-eq}), $\omega_l$, are in general complex with a negative imaginary part, i.e., $\text{Im}(\omega_l)<0$ ($l\in\{1,2,\dots,N\}$). To address the challenge posed by nonlinear root-finding problems, we employ the prescription of linearized Green's function, as detailed in Supplementary Note 6.

\subsection{Inverse participation ratio}
The inverse participation ratio (IPR) of the corresponding eigenvectors serves as a measure of their localization. The IPR is defined as~\cite{li2017mobility,hu2024emergent}:
\begin{equation}
\text{IPR}^{l}=\frac{\sum_{n=1}^{N}\left|p_{n}^{l}\right|^{4}}{\left(\sum_{n=1}^{N}\left|p_{n}^{l}\right|^{2}\right)^{2}}.
\end{equation}
From the definition, it is evident that $\text{IPR}^{l}\sim 1/N$ for an extended mode where $p_{n}^{l}\sim 1/\sqrt{N}$, while for a localized mode, where $p_{n}^{l}\sim \delta_{n,n_0}$, we have $\text{IPR}^{l}\sim 1$.

\subsection{Effective extinction efficiency}
 For a certain polarization, the local extinction efficiency $Q_n$ can be defined as~\cite{zundel2022green}
\begin{equation}
Q_{n}(\omega)=\frac{k}{\pi a^2\varepsilon_0}\text{Im}\left[A_{nn}(\omega)\right],
\end{equation}
where $A_{nn}$ represents the diagonal term of the matrix $\mathbf{A}$ with $\mathbf{A}=\left(\alpha^{-1}\mathbf{I}-\mathbf{M}\right)^{-1}$.
Here, $\mathbf{I}$ is an $N\times N$ identity matrix and $\mathbf{M}$ stands for an $N\times N$ interaction matrix whose elements are derived from the polarization-dependent Green's functions [Eqs.~(\ref{gf-xx}-\ref{gf-yz})]. For instance, for longitudinal polarization, $M_{nm}=\varepsilon_0^{-1}k^2G_{xx}(x_{nm})$ for $n\neq m$ and $M_{nn}=0$.

\section{Data availability}
The data that support the findings of this study are
available from the corresponding author upon request.

\section{Code availability}
The codes that support the findings in this study are available from the corresponding author upon request.

%

\section{Acknowledgments}
We gratefully acknowledge financial support from the Shenzhen Science and Technology Program (Grant Nos. RCYX20221008092848063, JCYJ20241202123733043, and JCYJ20241202123506009) and the National Natural Science Foundation of China (Grant Nos. 12574256 and 12447144).

\section{Author contributions}
Y.H. contributed to conceptualization, methodology,
data analysis, visualization, writing, and manuscript revision. K.Y. and W.H.X. contributed to the discussions of the results. X.C. led the project and contributed to project administration, supervision, writing review and editing, and funding acquisition. All authors reviewed and approved the final manuscript.

\section{Competing interests}
The authors declare no competing interests.


\end{document}